\documentstyle[sprocl]{article}

\begin{document}
\title{GETTING AT THE CP ANGLES} 
\author{{\bf A. I. Sanda} and {\bf Zhi-zhong Xing} }
\address{
Physics Department, Nagoya University, Chikusa-ku,
Nagoya 464-01, Japan}
\maketitle
\abstracts{
In anticipation for intensive experimental effort on $B$ physics,
theorists have introduced many ingenious ways to measure properties of 
the unitarity triangle. We review some of these methods. We will be critical
in the hope that, if there is any defect in them, it can be remedied either
theoretically or experimentally.
}

\section{Introduction}
\setcounter{equation}{0}

It has been known for more than thirty years that there exists $CP$
violation in the $K^0\Leftrightarrow \bar{K}^0$ transition
\cite{CP64}. This effect can be naturally interpreted by a non-trivial 
phase of the Kobayashi-Maskawa (KM) matrix \cite{KM73} for quark flavor mixing in
the standard electroweak model. So far, no other evidence for $CP$
violation has been unambiguously established.
Some intensive experimental efforts, such as the $B$
factory programs at KEK and at SLAC, are underway to search for large
signals of $CP$ asymmetries and to test the KM mechanism of $CP$
violation in the $B$-meson system. It is also expected that the study
of $B$ physics can provide a unique opportunity to discover new
physics beyond the standard model, in particular, that responsible for 
the origin of quark masses and $CP$ violation.

Unitarity of the $3\times 3$ KM matrix allows a geometrical
description of $CP$ violation in the complex plane, the so-called
unitarity triangle \cite{rss}. To meet various possible
measurements of $CP$ asymmetries at the forthcoming $B$ factories,
theorists have introduced many ingenious ways to determine properties
of the unitarity triangle. The aim of this talk is to review some
of those methods proposed in the past few years in a realistic manner.

To date, $K^0$-$\bar{K}^0$ mixing ($\Delta S=2$) and
$B^0_d$-$\bar{B}^0_d$ mixing ($\Delta B =2$) are the only second-order 
weak transition effects
that have been actually observed. If some new physics existed in the
$K$-meson or $B$-meson system, it is most likely to show itself in 
the $\langle K^0|{\cal H}|\bar{K}^0\rangle$ or 
$\langle B^0|{\cal H}|\bar{B}^0\rangle$
amplitude. New physics is less likely to compete
with the first-order weak interactions of the standard
model. The presence of new physics in $B^0$-$\bar{B}^0$
mixing will, in general, provide an additional weak phase to $CP$
asymmetries in neutral $B$ decays (due to the interplay of decay and
mixing). Thus the measurement of $CP$ violation and the KM unitarity
triangle at $B$ factories may serve as a useful approach towards
getting at the possible new physics beyond the standard model.

\section{Unitarity Triangle}
\setcounter{equation}{0}

One of the constraints among the KM matrix elements due to unitarity
is:
\begin{equation}
V^*_{ub}V_{ud} ~ + ~ V^*_{cb}V_{cd} ~ + ~ V^*_{tb}V_{td} \; = \; 0 \; .
%		(2.1)
\end{equation}
This relation corresponds to a triangle in the complex plane, the
well-known unitarity triangle \cite{rss} (see Fig. 1). 
%%%%%%%%%%%%%%%%%%%%%%%%%% Fig 1 %%%%%%%%%%%%%%%%%%%
\begin{figure}[t]
\begin{picture}(400,160)(-30,210)
\put(80,300){\line(1,0){150}}
\put(80,300.5){\line(1,0){150}}
\put(230,300){\vector(-1,0){75}}
\put(230,300.5){\vector(-1,0){75}}
\put(150,285.5){\makebox(0,0){$V^*_{cb}V_{cd}$}}
\put(80,300){\line(1,3){21.5}}
\put(80,300.5){\line(1,3){21.5}}
\put(80,299.5){\line(1,3){21.5}}
\put(80,300){\vector(1,3){10.75}}
\put(80,300.5){\vector(1,3){10.75}}
\put(80,299.5){\vector(1,3){10.75}}
\put(68,335){\makebox(0,0){$V^*_{ub}V_{ud}$}}
\put(230,300){\line(-2,1){128}}
\put(230,300.5){\line(-2,1){128}}
\put(101.5,364.5){\vector(2,-1){64}}
\put(101.5,365){\vector(2,-1){64}}
\put(176,343.5){\makebox(0,0){$V^*_{tb}V_{td}$}}
\put(95,310){\makebox(0,0){$\phi_3$}}
\put(188,311){\makebox(0,0){$\phi_1$}}
\put(108,350){\makebox(0,0){$\phi_2$}}
\end{picture}
\vspace{-3cm}
\caption{\small The unitarity triangle in the complex plane.}
\end{figure}
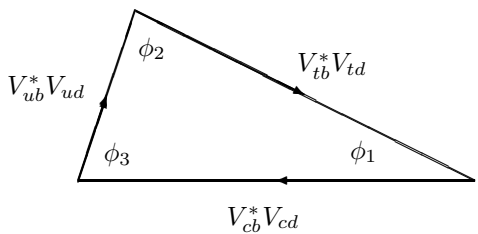
%%%%%%%%%%%%%%%%%%%%%%%%%%%%%%%%%%%%%%%%%%%%%%%%%%
Its three inner angles are denoted as
\begin{eqnarray}
\phi_1 & = & \arg \left (- \frac{V_{cb}^* V_{cd}}{V_{tb}^* V_{td}}
\right) \; , \nonumber \\
\phi_2 & = & \arg \left (- \frac{V_{tb}^* V_{td}}{V_{ub}^* V_{ud}} 
\right) \; , \nonumber \\
\phi_3 & = & \arg \left (- \frac{V_{ub}^* V_{ud}}{V_{cb}^* V_{cd}}
\right) \; .
%		(2.2)
\end{eqnarray}
If penguin and new physics effects are neglected, then $\phi_1$, $\phi_2$ and $\phi_3$
can be measured from $CP$ asymmetries in 
$B_d \rightarrow \psi K_S$, $B_d \rightarrow
\pi^+\pi^-$ and $B_s \rightarrow \rho^0K_S$, respectively
\cite{Sanda80}. 

Beyond the standard model, new physics may introduce an additional
$CP$-violating phase into $B^0_d$-$\bar{B}^0_d$ or $B^0_s$-$\bar{B}^0_s$ 
mixing. In this case, the phases {\it to be measured} from the above
decay modes (denoted by $\phi_{\psi K_S}$, $\phi_{\pi^+\pi^-}$ and
$\phi_{\rho^0 K_S}$, respectively) may deviate to some extent from the 
{\it geometrical} ones defined in (2.2). For the purpose of simplicity and
instruction, we only consider the kinds of new physics that do
not violate unitarity of the $3\times 3$ KM matrix \cite{NP}. Then 
$B^0_d$-$\bar{B}^0_d$ and $B^0_s$-$\bar{B}^0_s$ mixing phases can be
written as 
\begin{equation}
\left ( \frac{q}{p} \right )_{B_d} = \; \frac{V^*_{tb}
V_{td}}{V_{tb} V^*_{td}} ~ e^{2i\phi^d_{\rm NP}} \; , ~~~~~~ \left
( \frac{q}{p} \right )_{B_s} = \; \frac{V^*_{tb} V_{ts}}{V_{tb}
V^*_{ts}} ~ e^{2i\phi^s_{\rm NP}} \; ,
%		(2.3)
\end{equation}
where $\phi^d_{\rm NP}$ and $\phi^s_{\rm NP}$ denote the $CP$-violating
phases induced by new physics. Neglecting penguin effects and tiny
$CP$ violation in $K^0$-$\bar{K}^0$ mixing, we arrive at three angles
from three $CP$ asymmetries:
\begin{eqnarray}
{\rm Im} \left [ - \left (\frac{q}{p} \right )_{B_d}
\frac{V_{cb}V_{cs}^*}{V^*_{cb}V_{cs}} \right ] & = & \sin 2(\phi_1 -
\phi^d_{\rm NP}) \; , \nonumber \\
{\rm Im} \left [ + \left (\frac{q}{p}
\right )_{B_d} \frac{V_{ub}V^*_{ud}}{V^*_{ub}V_{ud}} \right ] & = &
\sin 2(\phi_2 + \phi^d_{\rm NP}) \; , \nonumber \\
{\rm Im} \left [ - \left ( \frac{q}{p}
\right )_{B_s} \frac{V_{ub}V^*_{ud}}{V^*_{ub}V_{ud}} \right ] & = &
\sin 2(\phi_3 - \phi^s_{\rm NP}) \; ,
%		(2.4)
\end{eqnarray}
where the sign ``$+$'' (or ``$-$'') comes from the $CP$ even (or odd)
final state. For simplicity in subsequent discussions, we define 
\begin{equation}
\phi_{\psi K_S} \; = \; \phi_1 - \phi^d_{\rm NP} \; , ~~~~
\phi_{\pi^+\pi^-} \; = \; \phi_2 + \phi^d_{\rm NP} \; , ~~~~
\phi_{\rho^0 K_S} \; = \; \phi_3 - \phi^s_{\rm NP} \; 
%		(2.5)
\end{equation}
as three {\it measurable} angles.
Two sides of the unitarity triangle, 
$|V^*_{ub}V_{ud}|$ and $|V^*_{cb}V_{cd}|$, have been
model-independently measured \cite{PDG96}, and will be improved in the future. 
A determination of the side
$|V^*_{tb}V_{td}|$ depends on the data of $B^0_d$-$\bar{B}^0_d$ mixing 
which might be affected by the presence of new physics.

Within the standard model, a detailed analysis of presently available
data yields the following constraints on three angles of the unitarity 
triangle \cite{Ali96}:
\begin{equation}
9^{\circ} \leq \phi_1 \leq 35^{\circ} \; , ~~~~~~~
45^{\circ} \leq \phi_2 \leq 148^{\circ} \; , ~~~~~~~
36^{\circ} \leq \phi_3 \leq 144^{\circ} \; .
%		(2.6)
\end{equation}
Note that $\phi_1 + \phi_2 + \phi_3 = 180^{\circ}$ is a natural
consequence of the above geometrical description. If a sum of 
$\phi_{\psi K_S}$, $\phi_{\pi^+\pi^-}$ and $\phi_{\rho^0 K_S}$
does not amount to $180^{\circ}$ at an acceptable
precision level, then $\phi^s_{\rm NP}\neq 0$.
However, an experimental confirmation of
the angle sum rule may not exclusively test the standard model
as it is not sensitive to the presence of $\phi^d_{\rm NP}$
\cite{Xing96}. We therefore stress that accurate
measurements of both angles and sides are necessary in order to fully
construct the unitarity triangle and pin down underlying new physics 
in the $B$ system.

\section{Penguin Pollution}
\setcounter{equation}{0}

The time-dependent rates of $B_d$ (or $B_s$) transitions to a $CP$
eigenstate $f$ can be given as \cite{Sanda80,Gronau89}
\begin{eqnarray}
\Gamma \left [\stackrel{(-)}{B^0}(t) \rightarrow f \right ] & \propto & 
e^{-\tau} \left [ \frac{1+ |\bar{\rho}^{~}_f|^2}{2}
~ \stackrel{(-)}{+} ~ \frac{1- |\bar{\rho}^{~}_f| ^2}{2} \cos (x \tau) \right
. \nonumber \\
& & ~~~~~~ \left . \stackrel{(+)}{-} ~ {\rm Im} \left ( \frac{q}{p} \bar{\rho}^{~}_f \right ) 
\sin ( x \tau ) \right ] \; , 
%		(3.1)
\end{eqnarray}
where $x=\Delta m/\Gamma$, $\tau=\Gamma t$, 
$\bar\rho^{~}_f = \langle f|{\cal H}|\bar{B}^0\rangle /\langle
f|{\cal H}|B^0\rangle$, and $q/p = e^{-2i\Phi_M}$ denotes the mixing phase. 
Then the $CP$ asymmetry between these two $CP$-conjugate processes
reads
\begin{eqnarray}
{\cal A}_f(t) & = & \frac{\Gamma (B^0(t) \rightarrow f) ~ - ~ 
{\Gamma} (\bar{B}^0(t) \rightarrow f)}{\Gamma (B^0(t) \rightarrow f) ~ + ~ 
\Gamma (\bar{B}^0(t) \rightarrow f)} \; \nonumber \\ \nonumber \\
& = & {\cal A}^{\rm c}_f \cos (x\tau) ~ + ~ {\cal A}^{\rm s}_f \sin
(x\tau) \; 
%		(3.2)
\end{eqnarray}
with 
\begin{equation}
{\cal A}^{\rm c}_f \; =\; \frac{1 - |\bar{\rho}^{~}_f|^2}{1 +
|\bar{\rho}^{~}_f|^2} \; , ~~~~~~~~
{\cal A}^{\rm s}_f \; =\; \frac{-2}{1 + |\bar{\rho}^{~}_f|^2} {\rm Im} \left (e^{-2i\Phi_M}
\bar{\rho}^{~}_f \right ) \; 
%		(3.3)
\end{equation}
denoting direct and indirect $CP$ asymmetries, respectively.

Due to the presence of penguin pollution, which may cause ${\cal A}^{\rm c}_f
\neq 0$, a $CP$ angle cannot be straightforwardly
extracted from the $CP$ asymmetry ${\cal A}^{\rm s}_f$. To see this
point more clearly, we decompose the decay amplitudes as
\begin{eqnarray}
\langle f|{\cal H}|B^0\rangle & = & e^{+i\Phi_1}e^{i\delta_1}A_1 ~ + ~ 
e^{+i\Phi_2}e^{i\delta_2}A_2 \; , \nonumber \\
\langle f|{\cal H}|\bar{B}^0\rangle & = & n^{~}_f \left [ e^{-i\Phi_1}e^{i\delta_1}A_1 
~ + ~ e^{-i\Phi_2}e^{i\delta_2}A_2 \right ] \; ,
%		(3.4)
\end{eqnarray}
where $n^{~}_f = \pm 1$ denotes the $CP$ parity of $f$,
$\Phi_1$ and $\Phi_2$ are weak phases, $\delta_1$ and $\delta_2$ 
are strong phases, $A_1$ and $A_2$ are magnitudes of hadronic matrix
elements and KM matrix elements. 
Then we obtain
\begin{equation}
\bar{\rho}^{~}_f \; = \; n^{~}_f e^{-2i\Phi_1} \left [ 1 ~ - ~ 2i \frac{A_2}{A_1}
\frac{\sin(\Phi_2-\Phi_1) e^{i(\delta_2-\delta_1)}}
{1 + e^{i(\Phi_2-\Phi_1)}e^{i(\delta_2-\delta_1)} A_2/A_1 } \right ]
\; .
%		(3.5)
\end{equation}
Without loss of generality one can take $A_1 > A_2$, i.e., the decay
amplitude is primarily governed by the $A_1$ component. 
Calculating ${\rm Im} (e^{-2i\Phi_M} \bar{\rho}^{~}_f)$ up to the leading terms of $A_2/A_1$, 
we arrive at
\begin{equation}
{\rm Im} \left (e^{-2i\Phi_M} \bar{\rho}^{~}_f \right ) 
\; \approx \; - n^{~}_f \left [ \sin 2 (\Phi_M + \Phi_1) ~ + ~
\Delta_f \right ] \; ,
%		(3.6)
\end{equation}
where
\begin{equation}
\Delta_f \; = \; 2 \frac{A_2}{A_1} \sin(\Phi_2-\Phi_1) \cos \left 
[ ( \delta_2 - \delta_1 ) - 2 (\Phi_M + \Phi_1) \right ] \; .
%		(3.7)
\end{equation}
Obviously the correction term $\Delta_f$ vanishes if 
$\Phi _2=\Phi_1$ or $A_2/A_1=0$. 

For illustration, we estimate the size of $A_2/A_1$ within the standard
model and link the phase combinations to the $CP$ angles for three
typical decay modes, as listed in Table 1. 
%%%%%%%%%%%%%%%%%% Table 1 %%%%%%%%%%%%%%%%%
\begin{table}[t]
\caption{Rough estimation of penguin pollution in the standard model
($\lambda \approx 0.22$).}
\vspace{-0.2cm}
\begin{center}
\begin{tabular}{lclcccc} \\ \hline\hline 
Example   & ~~~ & $A_2/A_1$ & ~~~ & $\sin (\Phi_2 - \Phi_1)$
& ~~~ & $\sin 2 (\Phi_M + \Phi_1)$  \\ \hline 
$B_d \rightarrow \psi K_S$	&&
$\sim \lambda^3$	&&
$\sim \lambda^2$	&&
$\sin (2\phi_1)$ \\ 
$B_d \rightarrow \pi^+\pi^-$	&&
$\sim \lambda$	&&
$\sin\phi_2$	&&
$\sin (2\phi_2)$ \\ 
$B_s \rightarrow \rho^0 K_S$	&&
$\sim 1$	&&
$\sin\phi_3$	&&
$\sin (2\phi_3)$ \\ 
\hline\hline
\end{tabular}
\end{center}
\end{table}
%%%%%%%%%%%%%%%%%%%%%%%%%%%%%%%%%%%%%%%%%%%%%
Clearly $\Delta_{\psi K_S}$ is safely negligible, while
$\Delta_{\pi^+\pi^-}$ and $\Delta_{\rho^0 K_S}$ may significantly
contaminate the extraction of $\phi_2$ and $\phi_3$ from 
${\cal A}^{\rm s}_{\pi^+\pi^-}$ and ${\cal A}^{\rm s}_{\rho^0 K_S}$,
respectively. As an example, the correlation between values of
$\sin (2\phi_2)$ and $\Delta_{\pi^+\pi^-}$ is illustrated in Fig. 2,
where $A_2/A_1 =0.2$ and $\delta_2 - \delta_1 =0$ have been typically
taken.
One can see that the penguin pollution in such decay 
channels have to be resolved \cite{SandaXing97,Gronau93}, 
in order to determine the relevant weak
angles reliably from their $CP$ asymmetries.
%%%%%%%%%%%%%%%%%%%%% Fig. 2 %%%%%%%%%%%%%%%%%
\begin{figure}
% GNUPLOT: LaTeX picture
\setlength{\unitlength}{0.240900pt}
\ifx\plotpoint\undefined\newsavebox{\plotpoint}\fi
\sbox{\plotpoint}{\rule[-0.500pt]{1.000pt}{1.000pt}}%
\begin{picture}(1200,990)(-110,-95)
\font\gnuplot=cmr10 at 10pt
\gnuplot
\sbox{\plotpoint}{\rule[-0.500pt]{1.000pt}{1.000pt}}
%\put(220.0,540.0){\rule[-0.500pt]{220.664pt}{1.000pt}}
%\put(678.0,113.0){\rule[-0.500pt]{1.000pt}{205.729pt}}
\put(220.0,113.0){\rule[-0.500pt]{4.818pt}{1.000pt}}
\put(198,113){\makebox(0,0)[r]{--1.0}}
\put(1116.0,113.0){\rule[-0.500pt]{4.818pt}{1.000pt}}
\put(220.0,198.0){\rule[-0.500pt]{4.818pt}{1.000pt}}
%\put(198,198){\makebox(0,0)[r]{-0.8}}
\put(1116.0,198.0){\rule[-0.500pt]{4.818pt}{1.000pt}}
\put(220.0,284.0){\rule[-0.500pt]{4.818pt}{1.000pt}}
\put(198,284){\makebox(0,0)[r]{--0.6}}
\put(1116.0,284.0){\rule[-0.500pt]{4.818pt}{1.000pt}}
\put(220.0,369.0){\rule[-0.500pt]{4.818pt}{1.000pt}}
%\put(198,369){\makebox(0,0)[r]{-0.4}}
\put(1116.0,369.0){\rule[-0.500pt]{4.818pt}{1.000pt}}
\put(220.0,455.0){\rule[-0.500pt]{4.818pt}{1.000pt}}
\put(198,455){\makebox(0,0)[r]{--0.2}}
\put(1116.0,455.0){\rule[-0.500pt]{4.818pt}{1.000pt}}
\put(220.0,540.0){\rule[-0.500pt]{4.818pt}{1.000pt}}
%\put(198,540){\makebox(0,0)[r]{0}}
\put(1116.0,540.0){\rule[-0.500pt]{4.818pt}{1.000pt}}
\put(220.0,625.0){\rule[-0.500pt]{4.818pt}{1.000pt}}
\put(198,625){\makebox(0,0)[r]{+0.2}}
\put(1116.0,625.0){\rule[-0.500pt]{4.818pt}{1.000pt}}
\put(220.0,711.0){\rule[-0.500pt]{4.818pt}{1.000pt}}
%\put(198,711){\makebox(0,0)[r]{0.4}}
\put(1116.0,711.0){\rule[-0.500pt]{4.818pt}{1.000pt}}
\put(220.0,796.0){\rule[-0.500pt]{4.818pt}{1.000pt}}
\put(198,796){\makebox(0,0)[r]{+0.6}}
\put(1116.0,796.0){\rule[-0.500pt]{4.818pt}{1.000pt}}
\put(220.0,882.0){\rule[-0.500pt]{4.818pt}{1.000pt}}
%\put(198,882){\makebox(0,0)[r]{0.8}}
\put(1116.0,882.0){\rule[-0.500pt]{4.818pt}{1.000pt}}
\put(220.0,967.0){\rule[-0.500pt]{4.818pt}{1.000pt}}
\put(198,967){\makebox(0,0)[r]{+1.0}}
\put(1116.0,967.0){\rule[-0.500pt]{4.818pt}{1.000pt}}
\put(220.0,113.0){\rule[-0.500pt]{1.000pt}{4.818pt}}
\put(220,68){\makebox(0,0){--1.0}}
\put(220.0,947.0){\rule[-0.500pt]{1.000pt}{4.818pt}}
\put(312.0,113.0){\rule[-0.500pt]{1.000pt}{4.818pt}}
%\put(312,68){\makebox(0,0){-0.8}}
\put(312.0,947.0){\rule[-0.500pt]{1.000pt}{4.818pt}}
\put(403.0,113.0){\rule[-0.500pt]{1.000pt}{4.818pt}}
\put(403,68){\makebox(0,0){--0.6}}
\put(403.0,947.0){\rule[-0.500pt]{1.000pt}{4.818pt}}
\put(495.0,113.0){\rule[-0.500pt]{1.000pt}{4.818pt}}
%\put(495,68){\makebox(0,0){-0.4}}
\put(495.0,947.0){\rule[-0.500pt]{1.000pt}{4.818pt}}
\put(586.0,113.0){\rule[-0.500pt]{1.000pt}{4.818pt}}
\put(586,68){\makebox(0,0){--0.2}}
\put(586.0,947.0){\rule[-0.500pt]{1.000pt}{4.818pt}}
\put(678.0,113.0){\rule[-0.500pt]{1.000pt}{4.818pt}}
%\put(678,68){\makebox(0,0){0}}
\put(678.0,947.0){\rule[-0.500pt]{1.000pt}{4.818pt}}
\put(770.0,113.0){\rule[-0.500pt]{1.000pt}{4.818pt}}
\put(770,68){\makebox(0,0){+0.2}}
\put(770.0,947.0){\rule[-0.500pt]{1.000pt}{4.818pt}}
\put(861.0,113.0){\rule[-0.500pt]{1.000pt}{4.818pt}}
%\put(861,68){\makebox(0,0){0.4}}
\put(861.0,947.0){\rule[-0.500pt]{1.000pt}{4.818pt}}
\put(953.0,113.0){\rule[-0.500pt]{1.000pt}{4.818pt}}
\put(953,68){\makebox(0,0){+0.6}}
\put(953.0,947.0){\rule[-0.500pt]{1.000pt}{4.818pt}}
\put(1044.0,113.0){\rule[-0.500pt]{1.000pt}{4.818pt}}
%\put(1044,68){\makebox(0,0){0.8}}
\put(1044.0,947.0){\rule[-0.500pt]{1.000pt}{4.818pt}}
\put(1136.0,113.0){\rule[-0.500pt]{1.000pt}{4.818pt}}
\put(1136,68){\makebox(0,0){+1.0}}
\put(1136.0,947.0){\rule[-0.500pt]{1.000pt}{4.818pt}}
\put(220.0,113.0){\rule[-0.500pt]{220.664pt}{1.000pt}}
\put(1136.0,113.0){\rule[-0.500pt]{1.000pt}{205.729pt}}
\put(220.0,967.0){\rule[-0.500pt]{220.664pt}{1.000pt}}
\put(40,540){\makebox(0,0){$\Delta_{\pi^+\pi^-}$}}
\put(678,-3){\makebox(0,0){$\sin (2\phi_2)$}}
\put(460,635){\makebox(0,0){\scriptsize $\phi_2 \in [135^{\circ}, 180^{\circ}]$}}
\put(900,635){\makebox(0,0){\scriptsize $\phi_2 \in [0^{\circ}, 45^{\circ}]$}}
\put(678,320){\makebox(0,){\scriptsize $\phi_2 \in [45^{\circ}, 135^{\circ}]$}}
\put(220.0,113.0){\rule[-0.500pt]{1.000pt}{205.729pt}}
%\put(678,967){\usebox{\plotpoint}}
%\put(678.0,113.0){\rule[-0.500pt]{1.000pt}{205.729pt}}
%\put(678,540){\usebox{\plotpoint}}
\multiput(678.00,541.86)(3.886,0.424){2}{\rule{6.050pt}{0.102pt}}
\multiput(678.00,537.92)(16.443,5.000){2}{\rule{3.025pt}{1.000pt}}
\multiput(707.00,546.84)(2.530,0.462){4}{\rule{4.917pt}{0.111pt}}
\multiput(707.00,542.92)(17.795,6.000){2}{\rule{2.458pt}{1.000pt}}
\multiput(735.00,552.86)(3.886,0.424){2}{\rule{6.050pt}{0.102pt}}
\multiput(735.00,548.92)(16.443,5.000){2}{\rule{3.025pt}{1.000pt}}
\multiput(764.00,557.86)(3.716,0.424){2}{\rule{5.850pt}{0.102pt}}
\multiput(764.00,553.92)(15.858,5.000){2}{\rule{2.925pt}{1.000pt}}
\put(792,560.92){\rule{6.745pt}{1.000pt}}
\multiput(792.00,558.92)(14.000,4.000){2}{\rule{3.373pt}{1.000pt}}
\multiput(820.00,566.86)(3.546,0.424){2}{\rule{5.650pt}{0.102pt}}
\multiput(820.00,562.92)(15.273,5.000){2}{\rule{2.825pt}{1.000pt}}
\put(847,569.92){\rule{6.263pt}{1.000pt}}
\multiput(847.00,567.92)(13.000,4.000){2}{\rule{3.132pt}{1.000pt}}
\put(873,573.42){\rule{6.263pt}{1.000pt}}
\multiput(873.00,571.92)(13.000,3.000){2}{\rule{3.132pt}{1.000pt}}
\put(899,576.42){\rule{5.782pt}{1.000pt}}
\multiput(899.00,574.92)(12.000,3.000){2}{\rule{2.891pt}{1.000pt}}
\put(923,579.42){\rule{5.782pt}{1.000pt}}
\multiput(923.00,577.92)(12.000,3.000){2}{\rule{2.891pt}{1.000pt}}
\put(947,581.92){\rule{5.541pt}{1.000pt}}
\multiput(947.00,580.92)(11.500,2.000){2}{\rule{2.770pt}{1.000pt}}
\put(970,583.42){\rule{5.300pt}{1.000pt}}
\multiput(970.00,582.92)(11.000,1.000){2}{\rule{2.650pt}{1.000pt}}
\put(1049,582.92){\rule{3.854pt}{1.000pt}}
\multiput(1049.00,583.92)(8.000,-2.000){2}{\rule{1.927pt}{1.000pt}}
\put(1065,580.92){\rule{3.373pt}{1.000pt}}
\multiput(1065.00,581.92)(7.000,-2.000){2}{\rule{1.686pt}{1.000pt}}
\put(1079,578.42){\rule{3.132pt}{1.000pt}}
\multiput(1079.00,579.92)(6.500,-3.000){2}{\rule{1.566pt}{1.000pt}}
\put(1092,574.92){\rule{2.891pt}{1.000pt}}
\multiput(1092.00,576.92)(6.000,-4.000){2}{\rule{1.445pt}{1.000pt}}
\put(1104,570.92){\rule{2.409pt}{1.000pt}}
\multiput(1104.00,572.92)(5.000,-4.000){2}{\rule{1.204pt}{1.000pt}}
\multiput(1114.00,568.71)(0.320,-0.424){2}{\rule{1.850pt}{0.102pt}}
\multiput(1114.00,568.92)(4.160,-5.000){2}{\rule{0.925pt}{1.000pt}}
\multiput(1122.00,563.69)(0.270,-0.462){4}{\rule{1.250pt}{0.111pt}}
\multiput(1122.00,563.92)(3.406,-6.000){2}{\rule{0.625pt}{1.000pt}}
\put(1127.92,554){\rule{1.000pt}{1.445pt}}
\multiput(1125.92,557.00)(4.000,-3.000){2}{\rule{1.000pt}{0.723pt}}
\put(1131.42,547){\rule{1.000pt}{1.686pt}}
\multiput(1129.92,550.50)(3.000,-3.500){2}{\rule{1.000pt}{0.843pt}}
\put(1133.42,540){\rule{1.000pt}{1.686pt}}
\multiput(1132.92,543.50)(1.000,-3.500){2}{\rule{1.000pt}{0.843pt}}
\put(1133.42,532){\rule{1.000pt}{1.927pt}}
\multiput(1133.92,536.00)(-1.000,-4.000){2}{\rule{1.000pt}{0.964pt}}
\put(1131.42,524){\rule{1.000pt}{1.927pt}}
\multiput(1132.92,528.00)(-3.000,-4.000){2}{\rule{1.000pt}{0.964pt}}
\put(1127.92,515){\rule{1.000pt}{2.168pt}}
\multiput(1129.92,519.50)(-4.000,-4.500){2}{\rule{1.000pt}{1.084pt}}
\multiput(1125.69,507.74)(-0.462,-0.579){4}{\rule{0.111pt}{1.750pt}}
\multiput(1125.92,511.37)(-6.000,-5.368){2}{\rule{1.000pt}{0.875pt}}
\multiput(1119.68,500.29)(-0.481,-0.470){8}{\rule{0.116pt}{1.375pt}}
\multiput(1119.92,503.15)(-8.000,-6.146){2}{\rule{1.000pt}{0.688pt}}
\multiput(1108.35,494.68)(-0.483,-0.485){10}{\rule{1.361pt}{0.117pt}}
\multiput(1111.17,494.92)(-7.175,-9.000){2}{\rule{0.681pt}{1.000pt}}
\multiput(1097.43,485.68)(-0.603,-0.485){10}{\rule{1.583pt}{0.117pt}}
\multiput(1100.71,485.92)(-8.714,-9.000){2}{\rule{0.792pt}{1.000pt}}
\multiput(1085.57,476.68)(-0.597,-0.487){12}{\rule{1.550pt}{0.117pt}}
\multiput(1088.78,476.92)(-9.783,-10.000){2}{\rule{0.775pt}{1.000pt}}
\multiput(1071.50,466.68)(-0.723,-0.485){10}{\rule{1.806pt}{0.117pt}}
\multiput(1075.25,466.92)(-10.252,-9.000){2}{\rule{0.903pt}{1.000pt}}
\multiput(1056.58,457.68)(-0.842,-0.485){10}{\rule{2.028pt}{0.117pt}}
\multiput(1060.79,457.92)(-11.791,-9.000){2}{\rule{1.014pt}{1.000pt}}
\multiput(1040.49,448.68)(-0.863,-0.487){12}{\rule{2.050pt}{0.117pt}}
\multiput(1044.75,448.92)(-13.745,-10.000){2}{\rule{1.025pt}{1.000pt}}
\multiput(1020.10,438.68)(-1.158,-0.481){8}{\rule{2.625pt}{0.116pt}}
\multiput(1025.55,438.92)(-13.552,-8.000){2}{\rule{1.313pt}{1.000pt}}
\multiput(1001.74,430.68)(-1.082,-0.485){10}{\rule{2.472pt}{0.117pt}}
\multiput(1006.87,430.92)(-14.869,-9.000){2}{\rule{1.236pt}{1.000pt}}
\multiput(979.55,421.68)(-1.364,-0.481){8}{\rule{3.000pt}{0.116pt}}
\multiput(985.77,421.92)(-15.773,-8.000){2}{\rule{1.500pt}{1.000pt}}
\multiput(955.32,413.69)(-1.665,-0.475){6}{\rule{3.536pt}{0.114pt}}
\multiput(962.66,413.92)(-15.661,-7.000){2}{\rule{1.768pt}{1.000pt}}
\multiput(931.73,406.69)(-1.746,-0.475){6}{\rule{3.679pt}{0.114pt}}
\multiput(939.36,406.92)(-16.365,-7.000){2}{\rule{1.839pt}{1.000pt}}
\multiput(907.73,399.69)(-1.746,-0.475){6}{\rule{3.679pt}{0.114pt}}
\multiput(915.36,399.92)(-16.365,-7.000){2}{\rule{1.839pt}{1.000pt}}
\multiput(879.97,392.69)(-2.325,-0.462){4}{\rule{4.583pt}{0.111pt}}
\multiput(889.49,392.92)(-16.487,-6.000){2}{\rule{2.292pt}{1.000pt}}
\multiput(850.38,386.71)(-3.377,-0.424){2}{\rule{5.450pt}{0.102pt}}
\multiput(861.69,386.92)(-14.688,-5.000){2}{\rule{2.725pt}{1.000pt}}
\put(820,379.92){\rule{6.504pt}{1.000pt}}
\multiput(833.50,381.92)(-13.500,-4.000){2}{\rule{3.252pt}{1.000pt}}
\put(792,375.92){\rule{6.745pt}{1.000pt}}
\multiput(806.00,377.92)(-14.000,-4.000){2}{\rule{3.373pt}{1.000pt}}
\put(764,372.42){\rule{6.745pt}{1.000pt}}
\multiput(778.00,373.92)(-14.000,-3.000){2}{\rule{3.373pt}{1.000pt}}
\put(735,369.92){\rule{6.986pt}{1.000pt}}
\multiput(749.50,370.92)(-14.500,-2.000){2}{\rule{3.493pt}{1.000pt}}
\put(707,368.42){\rule{6.745pt}{1.000pt}}
\multiput(721.00,368.92)(-14.000,-1.000){2}{\rule{3.373pt}{1.000pt}}
\put(678,367.42){\rule{6.986pt}{1.000pt}}
\multiput(692.50,367.92)(-14.500,-1.000){2}{\rule{3.493pt}{1.000pt}}
\put(649,367.42){\rule{6.986pt}{1.000pt}}
\multiput(663.50,366.92)(-14.500,1.000){2}{\rule{3.493pt}{1.000pt}}
\put(621,368.42){\rule{6.745pt}{1.000pt}}
\multiput(635.00,367.92)(-14.000,1.000){2}{\rule{3.373pt}{1.000pt}}
\put(592,369.92){\rule{6.986pt}{1.000pt}}
\multiput(606.50,368.92)(-14.500,2.000){2}{\rule{3.493pt}{1.000pt}}
\put(564,372.42){\rule{6.745pt}{1.000pt}}
\multiput(578.00,370.92)(-14.000,3.000){2}{\rule{3.373pt}{1.000pt}}
\put(536,375.92){\rule{6.745pt}{1.000pt}}
\multiput(550.00,373.92)(-14.000,4.000){2}{\rule{3.373pt}{1.000pt}}
\put(509,379.92){\rule{6.504pt}{1.000pt}}
\multiput(522.50,377.92)(-13.500,4.000){2}{\rule{3.252pt}{1.000pt}}
\multiput(486.38,385.86)(-3.377,0.424){2}{\rule{5.450pt}{0.102pt}}
\multiput(497.69,381.92)(-14.688,5.000){2}{\rule{2.725pt}{1.000pt}}
\multiput(463.97,390.84)(-2.325,0.462){4}{\rule{4.583pt}{0.111pt}}
\multiput(473.49,386.92)(-16.487,6.000){2}{\rule{2.292pt}{1.000pt}}
\multiput(441.73,396.84)(-1.746,0.475){6}{\rule{3.679pt}{0.114pt}}
\multiput(449.36,392.92)(-16.365,7.000){2}{\rule{1.839pt}{1.000pt}}
\multiput(417.73,403.84)(-1.746,0.475){6}{\rule{3.679pt}{0.114pt}}
\multiput(425.36,399.92)(-16.365,7.000){2}{\rule{1.839pt}{1.000pt}}
\multiput(394.32,410.84)(-1.665,0.475){6}{\rule{3.536pt}{0.114pt}}
\multiput(401.66,406.92)(-15.661,7.000){2}{\rule{1.768pt}{1.000pt}}
\multiput(373.55,417.83)(-1.364,0.481){8}{\rule{3.000pt}{0.116pt}}
\multiput(379.77,413.92)(-15.773,8.000){2}{\rule{1.500pt}{1.000pt}}
\multiput(353.74,425.83)(-1.082,0.485){10}{\rule{2.472pt}{0.117pt}}
\multiput(358.87,421.92)(-14.869,9.000){2}{\rule{1.236pt}{1.000pt}}
\multiput(333.10,434.83)(-1.158,0.481){8}{\rule{2.625pt}{0.116pt}}
\multiput(338.55,430.92)(-13.552,8.000){2}{\rule{1.313pt}{1.000pt}}
\multiput(316.49,442.83)(-0.863,0.487){12}{\rule{2.050pt}{0.117pt}}
\multiput(320.75,438.92)(-13.745,10.000){2}{\rule{1.025pt}{1.000pt}}
\multiput(298.58,452.83)(-0.842,0.485){10}{\rule{2.028pt}{0.117pt}}
\multiput(302.79,448.92)(-11.791,9.000){2}{\rule{1.014pt}{1.000pt}}
\multiput(283.50,461.83)(-0.723,0.485){10}{\rule{1.806pt}{0.117pt}}
\multiput(287.25,457.92)(-10.252,9.000){2}{\rule{0.903pt}{1.000pt}}
\multiput(270.57,470.83)(-0.597,0.487){12}{\rule{1.550pt}{0.117pt}}
\multiput(273.78,466.92)(-9.783,10.000){2}{\rule{0.775pt}{1.000pt}}
\multiput(257.43,480.83)(-0.603,0.485){10}{\rule{1.583pt}{0.117pt}}
\multiput(260.71,476.92)(-8.714,9.000){2}{\rule{0.792pt}{1.000pt}}
\multiput(246.35,489.83)(-0.483,0.485){10}{\rule{1.361pt}{0.117pt}}
\multiput(249.17,485.92)(-7.175,9.000){2}{\rule{0.681pt}{1.000pt}}
\multiput(239.68,497.00)(-0.481,0.470){8}{\rule{0.116pt}{1.375pt}}
\multiput(239.92,497.00)(-8.000,6.146){2}{\rule{1.000pt}{0.688pt}}
\multiput(231.69,506.00)(-0.462,0.579){4}{\rule{0.111pt}{1.750pt}}
\multiput(231.92,506.00)(-6.000,5.368){2}{\rule{1.000pt}{0.875pt}}
\put(223.92,515){\rule{1.000pt}{2.168pt}}
\multiput(225.92,515.00)(-4.000,4.500){2}{\rule{1.000pt}{1.084pt}}
\put(220.42,524){\rule{1.000pt}{1.927pt}}
\multiput(221.92,524.00)(-3.000,4.000){2}{\rule{1.000pt}{0.964pt}}
\put(218.42,532){\rule{1.000pt}{1.927pt}}
\multiput(218.92,532.00)(-1.000,4.000){2}{\rule{1.000pt}{0.964pt}}
\put(218.42,540){\rule{1.000pt}{1.686pt}}
\multiput(217.92,540.00)(1.000,3.500){2}{\rule{1.000pt}{0.843pt}}
\put(220.42,547){\rule{1.000pt}{1.686pt}}
\multiput(218.92,547.00)(3.000,3.500){2}{\rule{1.000pt}{0.843pt}}
\put(223.92,554){\rule{1.000pt}{1.445pt}}
\multiput(221.92,554.00)(4.000,3.000){2}{\rule{1.000pt}{0.723pt}}
\multiput(228.00,561.84)(0.270,0.462){4}{\rule{1.250pt}{0.111pt}}
\multiput(228.00,557.92)(3.406,6.000){2}{\rule{0.625pt}{1.000pt}}
\multiput(234.00,567.86)(0.320,0.424){2}{\rule{1.850pt}{0.102pt}}
\multiput(234.00,563.92)(4.160,5.000){2}{\rule{0.925pt}{1.000pt}}
\put(242,570.92){\rule{2.409pt}{1.000pt}}
\multiput(242.00,568.92)(5.000,4.000){2}{\rule{1.204pt}{1.000pt}}
\put(252,574.92){\rule{2.891pt}{1.000pt}}
\multiput(252.00,572.92)(6.000,4.000){2}{\rule{1.445pt}{1.000pt}}
\put(264,578.42){\rule{3.132pt}{1.000pt}}
\multiput(264.00,576.92)(6.500,3.000){2}{\rule{1.566pt}{1.000pt}}
\put(277,580.92){\rule{3.373pt}{1.000pt}}
\multiput(277.00,579.92)(7.000,2.000){2}{\rule{1.686pt}{1.000pt}}
\put(291,582.92){\rule{3.854pt}{1.000pt}}
\multiput(291.00,581.92)(8.000,2.000){2}{\rule{1.927pt}{1.000pt}}
\put(992.0,586.0){\rule[-0.500pt]{13.731pt}{1.000pt}}
\put(364,583.42){\rule{5.300pt}{1.000pt}}
\multiput(364.00,583.92)(11.000,-1.000){2}{\rule{2.650pt}{1.000pt}}
\put(386,581.92){\rule{5.541pt}{1.000pt}}
\multiput(386.00,582.92)(11.500,-2.000){2}{\rule{2.770pt}{1.000pt}}
\put(409,579.42){\rule{5.782pt}{1.000pt}}
\multiput(409.00,580.92)(12.000,-3.000){2}{\rule{2.891pt}{1.000pt}}
\put(433,576.42){\rule{5.782pt}{1.000pt}}
\multiput(433.00,577.92)(12.000,-3.000){2}{\rule{2.891pt}{1.000pt}}
\put(457,573.42){\rule{6.263pt}{1.000pt}}
\multiput(457.00,574.92)(13.000,-3.000){2}{\rule{3.132pt}{1.000pt}}
\put(483,569.92){\rule{6.263pt}{1.000pt}}
\multiput(483.00,571.92)(13.000,-4.000){2}{\rule{3.132pt}{1.000pt}}
\multiput(509.00,567.71)(3.546,-0.424){2}{\rule{5.650pt}{0.102pt}}
\multiput(509.00,567.92)(15.273,-5.000){2}{\rule{2.825pt}{1.000pt}}
\put(536,560.92){\rule{6.745pt}{1.000pt}}
\multiput(536.00,562.92)(14.000,-4.000){2}{\rule{3.373pt}{1.000pt}}
\multiput(564.00,558.71)(3.716,-0.424){2}{\rule{5.850pt}{0.102pt}}
\multiput(564.00,558.92)(15.858,-5.000){2}{\rule{2.925pt}{1.000pt}}
\multiput(592.00,553.71)(3.886,-0.424){2}{\rule{6.050pt}{0.102pt}}
\multiput(592.00,553.92)(16.443,-5.000){2}{\rule{3.025pt}{1.000pt}}
\multiput(621.00,548.69)(2.530,-0.462){4}{\rule{4.917pt}{0.111pt}}
\multiput(621.00,548.92)(17.795,-6.000){2}{\rule{2.458pt}{1.000pt}}
\multiput(649.00,542.71)(3.886,-0.424){2}{\rule{6.050pt}{0.102pt}}
\multiput(649.00,542.92)(16.443,-5.000){2}{\rule{3.025pt}{1.000pt}}
\put(307.0,586.0){\rule[-0.500pt]{13.731pt}{1.000pt}}
\end{picture}
\vspace{-.73cm}
%\vspace{0.15cm}
\caption{Illustrative plot for the correlation between $\Delta_{\pi^+\pi^-}$
and $\sin(2\phi_2)$.}
\end{figure}
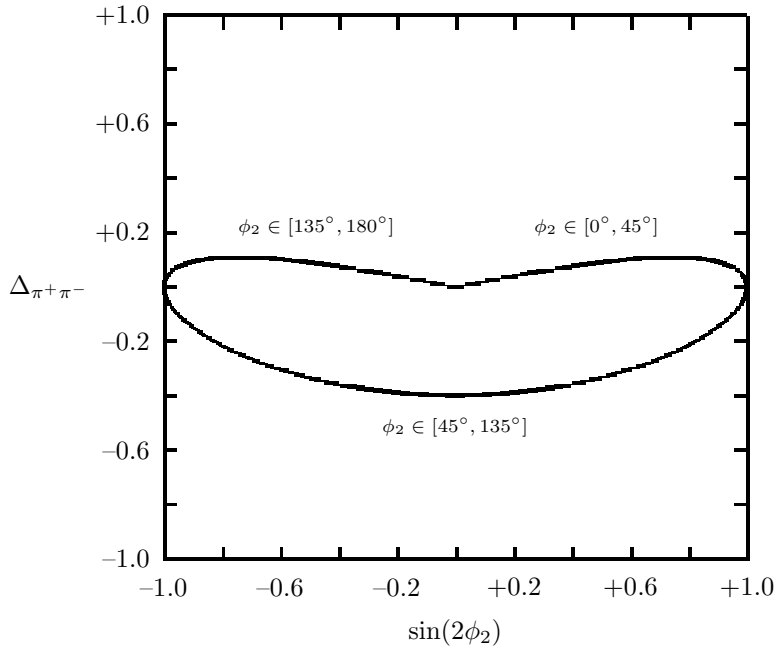
%%%%%%%%%%%%%%%%%%%%%%%%%%%%%%%%%%%%%%%%%%%%%%

\section{Cleanup of Penguin Pollution}
\setcounter{equation}{0}

To cleanly extract the $CP$ angle $\phi_2$ from $B_d \rightarrow \pi^+\pi^-$
or other charmless $B_d$ transitions, the relevant penguin and
tree-level effects should be disentangled. One can get around the
problem of penguin pollution by making use of isospin
relations. Subsequently we take two examples to illustrate this
method.

\underline{\it Example 1: $B\rightarrow \pi\pi$}.  
For $B\rightarrow 2\pi$ decays, an isospin analysis of 
$B^0_d\rightarrow \pi^+\pi^-$, $B^0_d\rightarrow \pi^0\pi^0$,
$B^+_u\rightarrow \pi^+\pi^0$ and their charge-conjugate processes is
possible \cite{Gronau90}. The amplitudes of these decay modes
are related by isospin symmetry as follows:
\begin{eqnarray}
& A^{+-} \; = \; \sqrt{2} (A_2 - A_0) \; , ~~~~~
A^{00} \; = \; 2A_2 + A_0 \; , ~~~~~
A^{+0} \; = \; 3A_2 \; , \nonumber \\
& \bar{A}^{+-} \; = \; \sqrt{2} (\bar{A}_2 - \bar{A}_0 ) \; , ~~~~~
\bar{A}^{00} \; =\; 2 \bar{A}_2 + \bar{A}_0 \; , ~~~~~
\bar{A}^{-0} \; =\; 3 \bar{A}_2 \; ,
%		(4.1)
\end{eqnarray}
where $A_I$ and $\bar{A}_I$ ($I=1, 0$) are isospin amplitudes. 
There exist two triangular relations in the complex plane:
\begin{equation}
A^{+-} + \sqrt{2} A^{00} \; = \; \sqrt{2} A^{+0} \; , ~~~~~~
\bar{A}^{+-} + \sqrt{2} \bar{A}^{00} \; =\; \sqrt{2} \bar{A}^{-0} \; .
%		(4.2)
\end{equation}
In terms of isospin amplitudes, $\bar{\rho}^{~}_{\pi^+\pi^-}$ and
$\bar{\rho}^{~}_{\pi^0\pi^0}$ read
\begin{equation}
\bar{\rho}^{~}_{\pi^+\pi^-} \; = \; \frac{\bar{A}^{+-}}{A^{+-}} \; =\; 
\frac{\bar{A}_2}{A_2} \frac{1 - \bar{z}}{1 - z} \; , ~~~~
\bar{\rho}^{~}_{\pi^0\pi^0} \; = \; \frac{\bar{A}^{00}}{A^{00}} \; =\; 
\frac{\bar{A}_2}{A_2} \frac{2 + \bar{z}}{2 + z} \; ,
%		(4.3)
\end{equation}
where $z = A_0/A_2$ and  ${\bar z} = {\bar A_0}/{\bar A_2}$ can be
solved from the isospin triangles in (4.2). Neglecting the electroweak 
penguin effect, which is expected to be small enough for our present
purpose \cite{Deshpande95,Gronau95}, we get
$\bar{A}_2/A_2 = (V_{ub}V^*_{ud})/(V^*_{ub}V_{ud})$
as a pure weak phase from the tree-level quark diagrams. It turns out that
\begin{eqnarray}
{\rm Im} \left ( e^{-2i\Phi_M} \bar{\rho}^{~}_{\pi^+\pi^-} \right ) & = &
{\rm Im} \left ( e^{2i\phi_{\pi^+\pi^-}} \frac{1 - \bar{z}}{1 - z} \right ) \; , \nonumber 
\\
{\rm Im} \left ( e^{-2i\Phi_M} \bar{\rho}^{~}_{\pi^0\pi^0} \right ) & = &
{\rm Im} \left ( e^{2i\phi_{\pi^+\pi^-}} \frac{2 + \bar{z}}{2 + z} \right ) \; ,
%		(4.4)
\end{eqnarray}
where $\phi_{\pi^+\pi^-} = \phi_2 + \phi^d_{\rm NP}$, given in (2.5).
Thus the $CP$ angle $\phi_2$ can be sorted out in spite of the penguin
pollution, if there is no new physics contribution to
$B^0_d$-$\bar{B}^0_d$ mixing (i.e., $\phi^d_{\rm NP} =0$).

The feasibility of this isospin method to extract $\phi_{\pi^+\pi^-}$ relies on
the observation of $B^0_d \rightarrow \pi^0\pi^0$ and
$\bar{B}^0_d\rightarrow \pi^0\pi^0$, whose branching ratios are
expected to be very small ($\sim 10^{-6}$) due to color suppression.
If the branching ratio of $B^0_d \rightarrow \pi^0\pi^0$ were too small
(e.g., $\leq 10^{-7}$) to be measured, one could take $A^{00} \approx
\bar{A}^{00} \approx 0$ as an effective approximation. In this limit,
$z \approx \bar{z} \approx -2$ or $\bar{\rho}^{~}_{\pi^+\pi^-} \approx 1$ would 
hold, leading to ${\rm Im} (e^{-2i\Phi_M} \bar{\rho}^{~}_{\pi^+\pi^-} )
\approx \sin (2\phi_{\pi^+\pi^-})$ with little penguin contamination. 

In practice, a model-dependent constraint on the branching ratio of 
$B^0_d \rightarrow \pi^0\pi^0$ may be obtained from the above isospin
analysis:
\begin{equation}
{\cal B} (B^0_d\rightarrow \pi^0\pi^0) \; \geq \; \frac{1}{2} \left
( \sqrt{2} - \sqrt{\frac{{\cal B} (B^0_d \rightarrow \pi^+\pi^-)}{{\cal B}
(B^+_u \rightarrow \pi^+\pi^0)}} \right )^2 {\cal B} (B^+_u\rightarrow 
\pi^+\pi^0) \; ,
%		(4.5)
\end{equation}
once the modes $B^+_u\rightarrow \pi^+\pi^0$ and $B^0_d\rightarrow
\pi^+\pi^-$ are reliably measured. This lower bound can be used to
check those model-dependent calculations for ${\cal B}
(B^0_d\rightarrow \pi^0\pi^0)$, and to examine if the approximation 
${\cal B} (B^0_d\rightarrow \pi^0\pi^0) \ll {\cal B} (B^0_d\rightarrow 
\pi^+\pi^-)$ (or $A^{00} \approx \bar{A}^{00} \approx 0$) is
acceptable in reality. As the $B^0_d\rightarrow \pi^0\pi^0$ transition
is of crucial importance in determining $\phi_2$, it is worthwhile to
make all possible experimental efforts to detect it.

\underline{\it Example 2: $B \rightarrow \rho \pi$}. For $B\rightarrow 
\rho \pi$ transitions, the final states include $I=0, 1$ and 2 isospin 
configurations and thus are more complicated than those in
$B\rightarrow \pi\pi$ decays. A detailed isospin analysis of
$B^+_u\rightarrow \rho^+\pi^0$, $B^+_u\rightarrow \rho^0\pi^+$,
$B^0_d\rightarrow \rho^+\pi^-$, $B^0_d\rightarrow \rho^-\pi^+$ and
$B^0_d\rightarrow \rho^0\pi^0$ has been made by Lipkin {\it et al}
\cite{Lipkin91}. For 
simplicity, one may distinguish between the tree-level ($T$) and
penguin ($P$) contributions to each overall decay amplitude, as the
penguin effect is purely of $I=1/2$ transition. Then five decay
amplitudes can be written as
\begin{eqnarray}
A^{+0} & = & (T^{+0} + 2P_1)/\sqrt{2} \; , \nonumber \\
A^{0+} & = & (T^{0+} - 2P_1)/\sqrt{2} \; , \nonumber \\
A^{+-} & = & T^{+-} + P_1 + P_0 \; , \nonumber \\
A^{-+} & = & T^{-+} - P_1 + P_0 \; , \nonumber \\
A^{00} & = & (T^{+0} + T^{0+} - T^{+-} - T^{-+} - 2P_0)/2 \; ,
%		(4.6)
\end{eqnarray}
in which $T^{00}$ is eliminated in terms of $T^{+0}$, $T^{0+}$,
$T^{+-}$ and $T^{-+}$ due to isospin constraints. For the
charge-conjuate channels $B^-_u\rightarrow \rho^-\pi^0$,
$B^-_u\rightarrow \rho^0\pi^-$, $\bar{B}^0_d\rightarrow \rho^-\pi^+$,
$\bar{B}^0_d\rightarrow \rho^+\pi^-$ and $\bar{B}^0_d\rightarrow
\rho^0\pi^0$, we define their amplitudes as $\bar{A}^{-0}$,
$\bar{A}^{0-}$, $\bar{A}^{-+}$, $\bar{A}^{+-}$ and $\bar{A}^{00}$,
respectively. The corresponding tree-level and penguin amplitudes can
be denoted by $\bar{T}$ and $\bar{P}$, which differ from $T$ and $P$
only in the sign of the KM phases. Then we arrive at two pentagonal
relations:
\begin{eqnarray}
\sqrt{2} (A^{+0} + A^{0+} ) & = & A^{+-} + A^{-+} + 2 A^{00} \; ,
\nonumber \\
\sqrt{2} (\bar{A}^{-0} + \bar{A}^{0-} ) & = & \bar{A}^{-+} +
\bar{A}^{+-} + 2 \bar{A}^{00} \; ,
%		(4.7)
\end{eqnarray}
contrasting with the simple triangular relations of $B\rightarrow
\pi\pi$ given in (4.2).

Measurements of the above ten decay rates can, (at least) in
principle, determine the ten sides of both pentagons in
(4.7). Furthermore, observation of $CP$ asymmetries in $B^0_d$ vs
$\bar{B}^0_d \rightarrow \rho^+\pi^-$, $\rho^-\pi^+$ and $\rho^0\pi^0$ 
will allow one to determine the following three quantities (here again 
the electroweak penguin effects are negligible \cite{Du97}):
\begin{equation}
{\rm Im} \left (e^{2i\phi_{\pi^+\pi^-}} \frac{\bar{A}^{+-}}{A^{+-}} \right ) \; ,
~~~
{\rm Im} \left (e^{2i\phi_{\pi^+\pi^-}} \frac{\bar{A}^{-+}}{A^{-+}} \right ) \; , 
~~~
{\rm Im} \left (e^{2i\phi_{\pi^+\pi^-}} \frac{\bar{A}^{00}}{A^{00}} \right ) \; .
%		(4.8)
\end{equation}
Thus the $CP$ angle $\phi_{\pi^+\pi^-}$ can be extracted without the (strong) penguin
pollution. This method, however, may be plagued with multiple discrete ambiguities
\cite{g91,Snyder93}. To avoid this drawback Quinn and Snyder have considered
a maximum-likelihood fit of the
parameters to the full Dalitz plot distribution, which is possible to
successfully extract $\phi_{\pi^+\pi^-}$ and other parameters with as few as
$10^3$ Monte-Carlo-generated events \cite{Snyder93}. Of course, such
an estimate relies on the assumption that $B\rightarrow 3\pi$ events
are fully dominated by $B\rightarrow \rho\pi$ \cite{Tanaka97}.
If the detector efficiency and
background effects are taken into account, for the practical purpose,
perhaps about $10^4$ $B\rightarrow \rho\pi$ events are required. 
This implies that we need to accumulate as many as $10^9$ $B\bar{B}$ events, 
which may be beyond what can be achieved in the first-round experiments of a $B$ factory.

\section{Extraction of $\phi_1 -\phi_2$ and $\phi_3$}
\setcounter{equation}{0}

Some neutral $B$ decays to hadronic non-$CP$ eigenstates can also be
used to extract the $CP$ angles. In particular, the angle difference
$\phi_1 -\phi_2$ is associated with the $CP$-violating quantity in
$B^0_d$ vs $\bar{B}^0_d \rightarrow D^{(*)0} K_S$ or
$\bar{D}^{(*)0} K_S$; while
$\phi_3$ is associated with the $CP$-violating quantity in
$B^0_s$ vs $\bar{B}^0_s \rightarrow D^{(*)0} \phi$ or $\bar{D}^{(*)0}
\phi$. In addition, $\phi_3$ is responsible for $CP$ violation
in the transitions $B^0_s$ vs
$\bar{B}^0_s \rightarrow D^{(*) \pm}_s K^{(*) \mp}$. 
Subsequently 
we shall analyze these three types of decays in some detail.

\underline{(a) ~ $B^0_d$ vs $\bar{B}^0_d \rightarrow D^{(*)0} K_S$ or
$\bar{D}^{(*)0} K_S$}. Such decay modes can only occur through the
tree-level quark transitions $b\rightarrow u \bar{c} s$ and $b
\rightarrow c \bar{u} s$; thus each decay amplitude involves only a
single weak phase \cite{Sanda88}, i.e., $\arg (V_{ub}V_{cs}^*)$ or $\arg
(V_{cb}V_{us}^*)$. Neglecting tiny $CP$ violation from $K^0$-$\bar{K}^0$ 
mixing in the final state, we parametrize the four transition amplitudes as
follows:
\begin{eqnarray}
\langle D^{(*)0} K_S |{\cal H}| B^0_d\rangle & = & (V^*_{ub} V_{cs}) 
A_1 e^{i\delta_1} \; , \nonumber \\
\langle \bar{D}^{(*)0} K_S |{\cal H}| \bar{B}^0_d\rangle & = & (V_{ub} 
V_{cs}^*) A_1 e^{i\delta_1} \; , \nonumber \\
\langle \bar{D}^{(*)0} K_S |{\cal H}| B^0_d\rangle & = & (V^*_{cb}
V_{us}) A_2 e^{i\delta_2} \; , \nonumber \\
\langle D^{(*)0} K_S |{\cal H}| \bar{B}^0_d\rangle & = & (V_{cb} 
V_{us}^*) A_2 e^{i\delta_2} \; , 
%		(5.1)
\end{eqnarray}
where $A_i$ and $\delta_i$ ($i=1,2$) are the real hadronic
matrix element and the corresponding strong phase. A time-dependent 
measurement of the above decay modes will allow one to determine two
quantities \cite{Bigi91}:
\begin{eqnarray}
{\rm Im} \left ( e^{-2i\Phi_M} \bar{\rho}_{D^{(*)0} K_S} \right ) &
\approx & \left | \bar{\rho}_{D^{(*)0} K_S} \right | \sin (\phi_{\psi K_S}
- \phi_{\pi^+\pi^-} - \Delta \delta) \; , \nonumber \\
{\rm Im} \left ( e^{-2i\Phi_M} \bar{\rho}_{\bar{D}^{(*)0} K_S} 
\right ) & \approx & \left | \bar{\rho}_{\bar D^{(*)0} K_S} \right |
\sin (\phi_{\psi K_S} - \phi_{\pi^+\pi^-} + \Delta \delta) \; , 
%		(5.2)
\end{eqnarray}
where $\Delta \delta = \delta_2 - \delta_1$, and $\phi_{\psi K_S} =
\phi_1 - \phi^d_{\rm NP}$ and $\phi_{\pi^+\pi^-} = \phi_2 + \phi^d_{\rm
NP}$ have been given in (2.5). 
Since  $|\bar{\rho}_{D^{(*)0} K_S}|$ and $|\bar\rho_{\bar{D}^{(*)0}
K_S}|$ can also be determined from the time-dependent measurement (see
(3.1) for illustration), one may extract both $\phi_{\psi K_S} - \phi_{\pi^+\pi^-}$ 
and $\Delta \delta$ from (5.2). 
Note that the strong phase shifts $\Delta \delta (D^0 K_S)$ and
$\Delta \delta (D^{*0}K_S)$ are expected to be different in general,
thus it is possible to resolve the discrete
ambiguity associated with the determination of $\phi_{\psi K_S}
-\phi_{\pi^+\pi^-}$. In the absence of new physics (i.e., $\phi^d_{\rm
NP} = 0$), this weak phase difference amounts to $\phi_1 -
\phi_2$.

The feasibility of this method depends mainly on branching ratios
of relevant channels and the detection efficiency (in particular, for
reconstructing the final-state neutral $D$ mesons). A rough estimate
yields ${\cal B} (B^0_d \rightarrow D^{(*)0} K_S) \sim 10^{-5}$ 
and ${\cal B} (B^0_d \rightarrow \bar{D}^{(*)0} K_S) \sim 10^{-4}$.
If this estimation is true, then the method under discussion is
unlikely to work in the first-round experiments of $B$ factories.

As we have shown before, $\phi_1$ and $\phi_2$ will be separately
determined from some $B_d$ decays to $CP$ eigenstates like $\psi K_S$
and $\pi^+\pi^-$. Thus a comparison between the angle difference
$\phi_2 - \phi_1$ obtained from such measurements and that obtained
from (5.2) will be helpful in order to check the self-consistency of
the standard model predictions. 

\underline{(b) ~ $B^0_s$ vs $\bar{B}^0_s \rightarrow D^{(*) \pm}_s
K^{(*) \mp}$}. These decay modes are also governed by the tree-level
quark transitions $b\rightarrow u \bar{c} s$ and $b\rightarrow c
\bar{u} s$, hence they can be analyzed in a similar way as done for
$B^0_d$ vs $\bar{B}^0_d \rightarrow D^{(*)0} K_S$ or $\bar{D}^{(*)0}
K_S$. We find that the $CP$-violating quantities
in $B^0_s$ vs $\bar{B}^0_s \rightarrow D^{(*) \pm}_s K^{(*) \mp}$ are
associated with the $CP$ angle $\phi_3$; i.e.,
\begin{eqnarray}
{\rm Im} \left ( e^{-2i\Phi_M} \bar{\rho}_{D^{(*)+}_s K^{(*)-}}
\right ) & \approx & \left | \bar{\rho}_{D^{(*)+}_s K^{(*)-}} \right |
\sin ( 2\phi^s_{\rm NP} - \phi_3 + \Delta \delta) \; , \nonumber \\
{\rm Im} \left ( e^{-2i\Phi_M} \bar{\rho}_{\bar{D}^{(*)-}_s K^{(*)+}}
\right ) & \approx & \left | \bar{\rho}_{\bar{D}^{(*)-}_s K^{(*)+}}
\right | \sin (2\phi^s_{\rm NP} - \phi_3  - \Delta \delta) \; , 
%		(5.3)
\end{eqnarray}
where $\phi^s_{\rm NP}$ denotes the $CP$ phase from new physics in
$B^0_s$-$\bar{B}^0_s$ mixing (see (2.3) for illustration), and
$\Delta \delta$ stands for the strong phase difference.
We see that $\phi_3 -2\phi^s_{\rm NP}$ can be extracted from
(5.3). The feasibility of this method has been discussed in deatail by 
Aleksan, Dunietz and Kayser \cite{Aleksan92}. It might suffer
from the rapid rate of $B^0_s$-$\bar{B}^0_s$ oscillation during the
time-dependent measurement of relevant decay modes. As $B_s$ mesons
cannot be produced at the KEK and SLAC $B$ factories, more time is
needed to realize the above method at high-luminosity hadron
machines.

\underline{(c) ~ $B^0_s$ vs $\bar{B}^0_s \rightarrow D^{(*)0} \phi$ or 
$\bar{D}^{(*)0}\phi$}. The analysis of these decay modes is 
the same as that of $B^0_s$ vs $\bar{B}^0_s \rightarrow D^{(*)\pm}_s
K^{(*)\mp}$. The result similar to (5.3) can be obtained \cite{Dunietz95}, allowing one
to extract the $CP$ angle $\phi_3 -2\phi^d_{\rm NP}$. 
In comparison between methods (b) and (c),
we find that the former is more promising in practice, because the
relevant (color-favored) transitions have larger branching ratios and
the final-state (charged) particles are easier to detect.

\section{$B \rightarrow (D^0, \bar{D}^0, D_{1,2}) + X$ and $\phi_3$}
\setcounter{equation}{0}

There is a variety of two-body $B$ decays involving $D^0$,
$\bar{D}^0$ or $D_{1,2}$ in the final states, where $D_{1,2} \equiv
(D^0 \pm \bar{D}^0)/\sqrt{2}$  denotes the $CP$ eigenstates
of neutral $D$ mesons. Such decay modes are interesting as 
they can be used to determine the $CP$ angles \cite{Sanda88,Gronau91}. For
example, the $CP$-violating quantities in $B_d \rightarrow (D^0,
\bar{D}^0, D_{1,2}) + K_S$ and $B_s \rightarrow (D^0, \bar{D}^0,
D_{1,2}) + \phi$ are associated with $\phi_1 -\phi_2$ and $\phi_3$,
respectively, as discussed above. In the following we shall
concentrate on the charged $B$ decays $B^{\pm}_u \rightarrow (D^0,
\bar{D}^0, D_{1,2}) + K^{\pm}$ to outline the main feature of such
transitions as well as the possibility to extract $\phi_3$.

The decay modes $B^{\pm}_u \rightarrow (D^0, \bar{D}^0, D_{1,2}) +
K^{\pm}$ occur only through the tree-level quark processes
$b\rightarrow u \bar{c} s$ and $b \rightarrow c \bar{u} s$. Negelcting 
tiny $D^0$-$\bar{D}^0$ mixing in the final states $D_{1,2}K^{\pm}$, 
one can parametrize all six transition amplitudes in the Wolfenstein
phase convention:
\begin{eqnarray}
A(D^0K^+) & \equiv & \langle D^0 K^{+}|{\cal H}|B^+_u\rangle \;
=\; |V_{ub}V_{cs}| A_a e^{i(\delta_a + \phi_3)} \; , \nonumber \\
A(\bar{D}^0 K^-) & \equiv & \langle \bar{D}^0 K^{-}|{\cal H}|B^-_u\rangle \;
=\; |V_{ub}V_{cs}| A_a e^{i(\delta_a - \phi_3)} \; ;
%		(6.1)
\end{eqnarray}
\begin{eqnarray}
A(\bar{D}^0 K^+) & \equiv & \langle \bar{D}^0 K^{+}|{\cal H}|B^+_u\rangle \;
=\; |V_{cb}V_{us}| A_b e^{i\delta_b} \; , \nonumber \\
A(D^0 K^-) & \equiv & \langle D^0 K^{-}|{\cal H}|B^-_u\rangle \;
=\; |V_{cb}V_{us}| A_b e^{i\delta_b} \; ;
%		(6.2)
\end{eqnarray}
and
\begin{eqnarray}
A(D_{1,2} K^+) \equiv \langle D_{1,2} K^{+}|{\cal H}|B^+_u \rangle 
=  \frac{1}{\sqrt{2}} \left [A(D^0 K^+) \pm A(\bar{D}^0 K^+) \right ] \; , \nonumber \\
A(D_{1,2} K^-) \equiv \langle D_{1,2} K^{-}|{\cal H}|B^-_u \rangle 
= \frac{1}{\sqrt{2}} \left [ A(D^0 K^-) \pm A(\bar{D}^0 K^-) \right
] \;  
, 
%		(6.3)
\end{eqnarray}
where $A_i$ and $\delta_i$ ($i=a,b$) are the real hadronic matrix
element and the strong phase, respectively. Clearly $A(D^0K^+) =
A(\bar{D}^0 K^-) e^{2i\phi_3}$ and $A(\bar{D}^0 K^+) = A(D^0 K^-)$
hold. In the factorization approximation, $A_a/A_b \sim a_1/a_2
\approx 0.26$, where $a_1$ and $a_2$ are the Bauer-Stech-Wirbel 
factors. One can see from (6.3) that $A(D^0 K^+), A(\bar{D}^0 K^+),
A(D_{1,2} K^+)$ and $A(D^0 K^-), A(\bar{D}^0 K^-), A(D_{1,2} K^-)$ 
form two correlated triangles in the
complex plane, as illustrated by Fig. 3. Since all sides of these two
triangles can be determined from the decay rates of $B^{\pm}_u
\rightarrow (D^0, \bar{D}^0, D_{1,2}) + K^{\pm}$, The $CP$ angle
$\phi_3$ is then resolved \cite{Gronau91}.
%%%%%%%%%%%%%%%%%%%%%%%%%% Fig 3 %%%%%%%%%%%%%%%%%%%
\begin{figure}[t]
\begin{picture}(400,160)(0,210)
\put(80,300){\line(1,0){150}}
\put(80,300.5){\line(1,0){150}}
\put(80,300){\vector(1,0){75}}
\put(80,300.5){\vector(1,0){75}}
\put(150,289){\makebox(0,0){\scriptsize $A (\bar{D}^0 K^+) = A (D^0 K^-)$}}
\put(80,300){\line(1,2){30}}
\put(80,300.5){\line(1,2){30}}
\put(80,299.5){\line(1,2){30}}
\put(80,300){\vector(1,2){15}}
\put(80,300.5){\vector(1,2){15}}
\put(80,299.5){\vector(1,2){15}}
\put(65,332){\makebox(0,0){\scriptsize $\sqrt{2}A (D_1 K^+)$}}
\put(230,300){\line(-2,1){120}}
\put(230,300.5){\line(-2,1){120}}
\put(230,300){\vector(-2,1){94}}
\put(230,300){\vector(-2,1){94}}
\put(142,361){\makebox(0,0){\scriptsize $A (D^0 K^+)$}}
\put(80,300){\line(2,1){200}}
\put(80,300.5){\line(2,1){200}}
\put(80,300){\vector(2,1){120}}
\put(80,300.5){\vector(2,1){120}}
\put(199,381){\makebox(0,0){\scriptsize $\sqrt{2} A (D_1 K^-)$}}
\put(230,300){\line(1,2){50}}
\put(230,300.5){\line(1,2){50}}
\put(230,299.5){\line(1,2){50}}
\put(230,300){\vector(1,2){25}}
\put(230,300.5){\vector(1,2){25}}
\put(230,299.5){\vector(1,2){25}}
\put(277,347){\makebox(0,0){\scriptsize $A (\bar{D}^0 K^-)$}}
\put(223,317){\makebox(0,0){$2\phi_3$}}
\end{picture}
\vspace{-3cm}
\caption{\small Triangular relations among decay amplitudes 
of $B^{\pm}_u \rightarrow (D^0, \bar{D}^0, D_1) + K^{\pm}$.}
\end{figure}
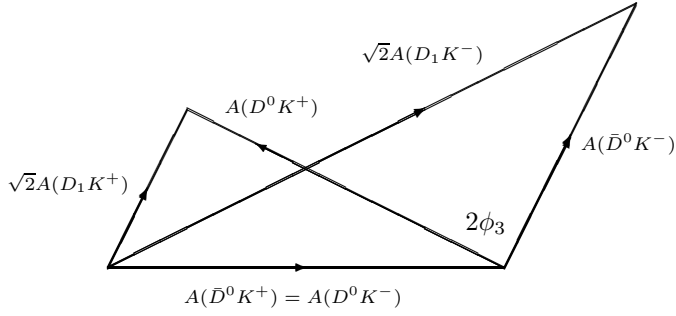
%%%%%%%%%%%%%%%%%%%%%%%%%%%%%%%%%%%%%%%%%%%%%%%%%%
Of course, the above discussion can be trivially extended to 
$B^0_d \rightarrow (D^0, \bar{D}^0, D_{1,2}) + K^{*0}$ 
and $B^{\pm}_c \rightarrow (D^0, \bar{D}^0, D_{1,2}) + D^{(*)\pm}_s$ decays.

In order to fully construct two correlated triangles in the complex
plane, their six sides should be comparable in magnitude. By use of the
factorization approximation, we estimate the ratio 
$|A(D^0 X)/ A(\bar{D}^0 X)|$ for three types of transitions mentioned
above ($X = K^{(*)+}$, $K^{*0}$ or $D^{(*)+}$) and list the rough
result in Table 2.
%%%%%%%%%%%%%%%%%% Table 2 %%%%%%%%%%%%%%%%%
\begin{table}[t]
\caption{Rough estimation of $|A(D^0 X)/A(\bar{D}^0 X)|$ for three types of 
$B$ decays.}
\vspace{-0.2cm}
\begin{center}
\begin{tabular}{lcc} \\ \hline\hline 
Transitions   & ~~~~~ & $|A(D^0 X)/A(\bar{D}^0 X)|$  \\ \hline 
$B^{\pm}_u \rightarrow (D^0, \bar{D}^0, D_{1,2}) + K^{(*)\pm}$	&&
$\sim \displaystyle \left | \frac{V_{ub}V_{cs}}{V_{cb}V_{us}} \frac{a_2}{a_1} \right | 
\sim 0.1$ \\ \\
$\stackrel{(-)}{B}$$^0_d \rightarrow (D^0, \bar{D}^0, D_{1,2}) + \stackrel{(-)}{K}$$^{*0}$	&&
$\sim \displaystyle \left |\frac{V_{ub}V_{cs}}{V_{cb}V_{us}} \frac{a_2}{a_2} \right | 
\sim 0.4$ \\ \\
$B^{\pm}_c \rightarrow (D^0, \bar{D}^0, D_{1,2}) + D^{(*)\pm}_s$	&&
$\sim \displaystyle \left |\frac{V_{ub}V_{cs}}{V_{cb}V_{us}} \frac{a_1}{a_2} \right | 
\sim 1.4$ \\ 
\hline\hline
\end{tabular}
\end{center}
\end{table}
%%%%%%%%%%%%%%%%%%%%%%%%%%%%%%%%%%%%%%%%%%%%%
For $B^{\pm}_u$ decays, one can see that the sides $A(D^0K^+)$ and
$A(\bar{D}^0 K^-)$ are too short in comparison with $A(\bar{D}^0 K^+)$ and
$A(D^0 K^-)$. Thus it will be very hard, if not even practically
impossible, to extract $\phi_3$ from these $B^{\pm}_u$ channels. 
Also, it is very difficult to identify $D^0$ in the suppressed decay
mode $B^+_u \rightarrow D^0 K^+$; e.g., if one identifies $D^0$ by
use of its leptonic decay, the same lepton from direct $B^+_u$ decay 
will swamp the signal \cite{Atwood97}.
A determination of $\phi_3$ from $B^{\pm}_c$ decays
\cite{Masetti92} might be possible at LHC-$B$.

To get around the above-mentioned problem associated with the
extraction of $\phi_3$ from $B^{\pm}_u \rightarrow (D^0, \bar{D}^0,
D_{1,2}) + K^{(*)\pm}$, Atwood, Dunietz and Soni \cite{Atwood97} have proposed the
measurement of $B^+_u \rightarrow D^0K^+ \rightarrow (K^-\pi^+)_{D^0}
K^+$, $B^+_u\rightarrow \bar{D}^0K^+ \rightarrow
(K^-\pi^+)_{\bar{D}^0} K^+$ and their charge-conjugate channels. The
point is that $D^0\rightarrow K^-\pi^+$ is Cabibbo-allowed while
$\bar{D}^0\rightarrow K^-\pi^+$ is doubly Cabibbo-suppressed, hence
the amplitudes $A [(K^-\pi^+)_{D^0}K^+]$ and 
$A[(K^-\pi^+)_{\bar{D}^0} K^+]$ become
comparable in magnitude:
\begin{equation}
\left | \frac{A [(K^-\pi^+)_{D^0} K^+]}{A [(K^-\pi^+)_{\bar{D}^0} K^+]} \right | \; 
\approx \; \left |\frac{V_{ub}V_{cs}}{V_{cb}V_{us}} \cdot \frac{a_2}{a_1}
\right | \cdot \left | \frac{\langle K^-\pi^+|{\cal H}|D^0\rangle }{\langle
K^-\pi^+|{\cal H}|\bar{D}^0\rangle} \right | 
\; \sim \; 1 \; .
%		(6.4)
\end{equation}
where ${\cal B}(\bar{D}^0\rightarrow K^-\pi^+)/{\cal B}(D^0\rightarrow 
K^-\pi^+) \approx 0.0077$ reported by CLEO \cite{CLEO94} has been used. Then the
overall amplitude of $B^+_u\rightarrow (K^-\pi^+)_D K^+$ involves
large interference between its two components. Experimentally one may also measure $B^+_u
\rightarrow (K^{*-}\pi^+)_D K^+$, $(K^-\rho^+)_D K^+$, $(K^-a^+_1)_D
K^+$ and their charge-conjugate processes, which have the same
weak interaction but different strong final-state interactions. This
allows the extraction of $\phi_3$ from four different decay rates, e.g.,
\small
\begin{eqnarray}
& & {\cal B}[B^+_u \rightarrow (K^-\pi^+)_D K^+] \; \nonumber \\
& = & {\cal B}(B^+_u \rightarrow D^0K^+) {\cal B}(D^0\rightarrow
K^-\pi^+) + {\cal B}(B^+_u \rightarrow \bar{D}^0 K^+) {\cal
B}(\bar{D}^0 \rightarrow K^-\pi^+) + \nonumber \\
&  & 2 \sqrt{{\cal B}(B^+_u \rightarrow D^0K^+) {\cal B}(D^0\rightarrow
K^-\pi^+) {\cal B}(B^+_u \rightarrow \bar{D}^0 K^+) {\cal
B}(\bar{D}^0 \rightarrow K^-\pi^+)} \nonumber \\
&  & \times \cos \left (\delta^{K\pi}_{ab} + \phi_3 \right ) \; ,
\nonumber \\
& & {\cal B}[B^-_u \rightarrow (K^+\pi^-)_D K^-] \; \nonumber \\
& = & {\cal B}(B^-_u \rightarrow \bar{D}^0K^-) {\cal B}(\bar{D}^0\rightarrow
K^+\pi^-) + {\cal B}(B^-_u \rightarrow D^0 K^-) {\cal
B}(D^0 \rightarrow K^+\pi^-) + \nonumber \\
&  & 2 \sqrt{{\cal B}(B^-_u \rightarrow \bar{D}^0K^-) {\cal B}(\bar{D}^0\rightarrow
K^+\pi^-) {\cal B}(B^-_u \rightarrow D^0 K^-) {\cal
B}(D^0 \rightarrow K^+\pi^-)} \nonumber \\
&  & \times \cos \left (\delta^{K\pi}_{ab} - \phi_3 \right ) \;
%		(6.5)
\end{eqnarray}
\normalsize
with $\delta^{K\pi}_{ab} = \delta^{K\pi}_a - \delta^{K\pi}_b$; and
those associated with $(K^*\pi)_D$ in the final state. 
Note that ${\cal B}(B^+_u \rightarrow D^0 K^+) = {\cal B}
(B^-_u\rightarrow \bar{D}^0 K^-)$, ${\cal B} (B^+_u \rightarrow
\bar{D}^0 K^+) = {\cal B}(B^-_u\rightarrow D^0 K^-)$, ${\cal B}
(D^0\rightarrow K^{(*)-}\pi^+) = {\cal B}(\bar{D}^0\rightarrow K^{(*)+}\pi^-)$
and ${\cal B}(D^0\rightarrow K^{(*)+}\pi^-) = {\cal B}(\bar{D}^0\rightarrow 
K^{(*)-}\pi^+)$. 
Therefore the four unknown 
quantities ${\cal B}(B^+_u\rightarrow D^0K^+)$,
$\delta^{K\pi}_{ab}$, $\delta^{K^*\pi}_{ab}$ and $\phi_3$ can all
be extracted from the measurement.
To resolve the discrete ambiguities in this
method, more decay modes of $D^0$ and $\bar{D}^0$ mesons should be
taken into account. It is expected that this method works in the
second-round experiments of a $B$ factory.

\section{SU(3) Analysis}
\setcounter{equation}{0}

Finally let us comment briely on the $SU(3)$ method of extracting the
$CP$ angles from $B$ decays, which has recently attracted a lot of
theorists' attention. Under flavor $SU(3)$ symmetry, amplitudes of a
variety of two-body mesonic decays, such as $B\rightarrow \pi\pi$,
$B\rightarrow K\pi$ and $B\rightarrow KK$, are related to one
another, allowing the possibility to determine the associated weak and 
strong phases. Intuitively the $SU(3)$ reduced matrix elements can be 
described in terms of a set of quark diagrams \cite{Zeppenfeld81}.
However, $SU(3)$ symmetry is expected to be broken by effects of order 
$20\%$ (e.g., $f_K/f_{\pi} \approx 1.2$), hence one has to introduce
appropriate $SU(3)$ breaking terms in explicit analyses. 
In most of $SU(3)$ analyses,
it is usually argued that the decay amplitudes via $W$-exchange, annihilation
and penguin annihilation diagrams are
formfactor suppressed and thus negligible as a good approximation.
This assumption can experimentally be checked if one detects the
expected suppression ${\cal B}(B^0_d \rightarrow K^+K^-)/{\cal
B}(B^0_d \rightarrow \pi^+\pi^-)$.

Many possibilities to extract three $CP$ angles from various $SU(3)$
relations of charmless $B$ decays have been proposed \cite{SU3}. For example,
the triangular relation \cite{GRL94}
\begin{equation}
\sqrt{2} A(B^+_u \rightarrow \pi^0K^+) + A(B^+_u\rightarrow
\pi^+K^0) = \sqrt{2} \left |\frac{V_{us}}{V_{ud}} \right |
\frac{f_K}{f_{\pi}} A(B^+_u \rightarrow \pi^+\pi^0)
\; 
%		(7.1)
\end{equation}
and its charge-conjugate relation can be used to determine the angle
$\phi_3$ in the standard model. 
Since several comprehensive reviews on these approaches have existed in the 
literature \cite{ReviewSU3}, we shall not go into any detail here.

Of course there are several sources of uncertainties associated with
the $SU(3)$ analysis (e.g., it is difficult to estimate the $SU(3)$
breaking effect in relevant strong phases). 
For this reason, this method cannot be used to look for new physics
effects; but it should be useful for self-consistency checks.

\section{Concluding Remarks}
\setcounter{equation}{0}

In this mini-review, we have discussed various ways to get at properties of the
unitarity triangle. We believe that $B^0$-$\bar{B}^0$ mixing is the
best bet to look for new physics. 
As $CP$ asymmetries in most of neutral $B$ decays involve
the interplay of decay and $B^0$-$\bar{B}^0$ mixing, some special
attention has been paid to the relationship between the {\it
geometrical} angles ($\phi_1$, $\phi_2$, $\phi_3$) and the {\it
measurable} ones ($\phi_{\psi K_S}$, $\phi_{\pi^+\pi^-}$,
$\phi_{\rho^0 K_S}$, etc). The latter may contain some information
about the underlying new physics in $B$ decays. Thus it is worthwhile 
to confront different methods of extracting $\phi_i$ with the forthcoming measurements 
at $B$-meson factories, in order to fully test the KM mechanism of
$CP$ violation and pin down possible new physics.

Eventually an accurate measurement of $|V_{ub}|$ and $|V_{cb}|$ will be
available to fix two sides of the unitarity triangle without much
interference from new physics. The element 
$|V_{td}|$ can be well determined (or constrained) from
observation of $K^+\to\pi^+\nu\bar\nu$. When the hadronic matrix element
associated with $\langle \bar{B}^0|{\cal H}|B^0 \rangle$ is measured
from more delicate lattice-QCD computation, there will be a more
reliable (and independent) constraint on $|V_{td}|$ from the data of
$B^0_d$-$\bar{B}^0_d$ mixing. These two measurements of $|V_{td}|$ may 
not agree to each other, however, if there is substantial new physics. 

It is quite clear that while a search for new physics in $B$ decays represents
a great experimental challenge, it might yield great reward. Also it
is a good time to think about more delicate $B$-physics experiments
beyond asymmetric $B$ factories.

{\bf Acknowledgments:} 
The work of A.I.S. was supported in part by the Grant-in-Aid for Scientific
Research on Priority Areas ({\it Physics of $CP$ Violation}) from the
Ministry of Education, Science and Culture of Japan.
Z.Z.X. thanks the Japan Society
for the Promotion of Science for its financial support.


\begin{thebibliography}{99}
\bibitem{SandaXing97} The angle $\phi_1$ may also be extracted from
$CP$ asymmetries in $B_d \rightarrow D^{(*)+}D^{(*)-}$, where the
penguin effects are not expected to be significant. See, e.g.,
A.I. Sanda and Z.Z. Xing, Phys. Rev. D {\bf 56}, 341 (1997).
\end{thebibliography}
\end{document}